*Article*

# Modelling of Quantum Dots with the Finite Element Method

**G.A. Mantashian[1,2]\*, P.A. Mantashyan [2,1]\*, D.B. Hayrapetyan [1,2]**

1   Department of General Physics and Quantum Nanostructures, Russian-Armenian University, Yerevan 0051, Armenia;

2   Institute of Chemical Physics after A.B. Nalbandyan, Yerevan 0014, Armenia;

\*   Correspondence: paytsar.mantashyan@rau.am; Tel.: +37493380236

**Abstract:** Considering the increasing number of experimental results in the manufacturing process of quantum dots (QDs) with different geometries, and the fact that most numerical methods that can be used to investigate quantum dots with non-trivial geometries require large computational capacities the finite element method (FEM) becomes an incredibly attractive tool for modeling semiconductor QDs. In the current article, the authors have used the FEM to obtain the first twenty-six probability densities and energy values for the following GaAs structures: rectangular, spherical, cylindrical, ellipsoidal, spheroidal, and conical QDs, quantum rings, nanotadpoles, and nanostars. The results of the numerical calculations were compared with the exact analytical solutions and a good deviation was obtained. The ground states energies dependence on the element size was obtained to find the optimal parameter for the investigated structures. The abovementioned calculation results were used to obtain valuable insight into the effects of the size quantization's dependence on the shape of the QDs. Additionally, the wavefunctions and energies of spherical CdSe/CdS quantum dots were obtained while taking into account the diffusion effects on the potential depth with the use of a piecewise Woods-Saxon potential. The diffusion of the effective mass and the dielectric permittivity is obtained with the use of a normal Woods-Saxon potential. A structure with a quasi type-II band alignment was obtained at the core size of $\approx 2.2nm$. This result is consistent with the experimental data.

## 1. Introduction

Semiconductor QDs have been a focus of condensed matter scientists for several decades. Their tunable properties mediated by the three-dimensional quantum confinement effect have led them to find application in numerous fields including but not limited to alternative energy [1-6], energy storage [7-11], sensing 12-17], quantum optics and photonics [18-25], quantum information [26-30], etc. Moreover, the newest advances in material science have led to the development of incredibly complex manufacturing and growth methods for QDs. These developments have led to the finding of more applications and the creation of a subclass of QDs: so-called QDs with non-trivial geometry. Examples of such structures vary from relatively simple like lens-shaped QDs [31], quantum rings [32-34], and nanorods [35-37] to more complex ones like nanotadpoles [38-40], nanostars [41-44], nanoscrews [45], tetrapods [46,47], nanodumbells [48,49], etc. Moreover, these structures are not limited to one material they can be grown like heterostructures with one region of the structure comprised of one material and the other region of another. Although experimental research has made great strides in the investigation of QDs with non-trivial geometry theoretical research has lagged. The reason for this lag is caused due to the fact that in most cases the attainment of electronic wavefunctions and energies for QD with non-trivial geometries is impossible. Some research was done in the framework of the envelope function approximation conjoined with the effective mass approximation for QDs with strongly oblate and strongly prolate geometries. In these cases, the geometrical adiabatic approximation can be used to obtain the eigenvalues and the eigenfunctions [50-54]. However, these cases are highly limited, and even relatively simple structures like conical QDs with comparable base radius and height cannot be investigated in such a way.

So naturally for cases where no analytical solution can be obtained the numerical methods come into play[55-57]. However, most of these methods like quantum chemistry methods require great computational capacities like clusters or supercomputers. Moreover, the calculation can take an extremely long time to complete. Although without a doubt the most accurate, the speed of these methods makes them non-flexible for cases when the research requires a variation of the geometrical parameters or a change of external fields.

With that taken into consideration, the FEM serves as a relatively low-cost computation-wise method that has good accuracy for obtaining a qualitative understanding of QDs with complex non-trivial geometry.

The FEM has been in extensive use by engineers for the last decade. The method has helped engineers make their designs safer and more cost-effective. The relatively fast speed of the simulations allows the method to be used in modeling the material properties and failure criteria for composite materials [54-61] both on a micro level and the macro level. For a three-dimensional system, the FEM is formulated in the following way the volume of the system is represented through finite elements. These elements are connected at nodal points located at the corners, sides, surface, or volume of the elements. These non-overlapping elements fill the volume of the system. In the end, a set of algebraic equations emerges from the formulation of a boundary value problem using the finite element approach. The technique makes domain-wide approximations of the unknown function. The simple equations that model these finite elements are then combined into a bigger system of equations that models the entire problem. The calculus of variations is used by the FEM to minimize an associated error function and then approximate a solution.

In comparison to the engineering field, the field of semiconductors has adapted this method relatively recently. However, the FEM has created opportunities for relatively high-accuracy theoretical research that would have been otherwise analytically impossible. The FEM calculations have paved the way for the investigation of electronic and excitonic states in core/shell and multilayer QDs both with the use of model potentials [62-64] and more direct simulational approaches [65-67].

In the following article [68], the FEM was used to theoretically investigate the electronic states in a core/shell pyramidal quantum dot with a GaAs core embedded in an AlGaAs matrix. The electronic states were investigated by analyzing the effects of the geometrical parameters and the external perturbations. The evaluation of the light absorption and relative refractive index changes, under different applied magnetic field configurations, was carried out.

In the next example [69], the properties of the electron states in the presence of a donor impurity in spherical sector-shaped quantum dots are obtained using the FEM within the framework of the effective mass approximation. The dependence of the spectrum on the radius and apical angle as well as in the position of the impurity is discussed. The comparison of the attained results with available experimental data for GaAs truncated-whisker-like quantum dots shows considerable agreement. The calculation has helped identify the lowest energy photoluminescence peak as donor-related.

Overall, the use of FEM simulations in recent years has grown considerably, showing accuracy and practicality. The method has helped investigated structures with complex geometries like nanotadpoles and core/shell structures [70-74] and even takes into account double structures [75].

This article aims to extensively investigate the applications of FEM for the simulation of the geometrically non-trivial QDs. The article has the following structure: Section 1: Introduction – This section contains the problem statement and the overview of the available scientific literature; Section 2: Materials and Methods – contains the methods for investigating both one material and core/shell quantum dots; Section 3: Results and Discussion – contains the detailed discussion of the probability density and the energy spectrum of the spherical, cylindrical, rectangular, conical, ellipsoidal QDs as well as quantum rings (QR) nanotadpoles, nanostars, core/shell; Section 4: Conclusion – this section contains the summary of the most important results.

## 2. Materials and Methods

The most investigated semiconductor material after silicon is gallium arsenide GaAs. It has a direct band structure and suitable parameters for many applications. As such for this article the GaAs is the most appropriate material. The material parameters that we are going to use in our calculations for GaAs are taken: $m_e^* = 0.067 m_0$ - electron effective mass, $m_h^* = 0.067 m_0$ - hole effective mass, $\varepsilon_s = 12.8$ - static dielectric susceptibility, $\varepsilon_\infty = 10.86$ - high-frequency dielectric constant, $E_R = 5.5\, meV$ - electron effective Rydberg energy, $a_{bohr} = 10.2\, nm$ - electron Bohr radius.

In the current article, the calculations have been carried out with the use of the commercial software called Wolfram Mathematica however, the obtained results and methods can be generalized for other software like MATLAB, COMSOL Multiphysics, etc. In any FEM calculation, the first step is the definition of the partial differential equation, in our case we are going to solve a three-dimensional one particle Schrodinger equation.

$$-\frac{\hbar^2}{2m^*}\left(\frac{\partial^2}{\partial x^2} + \frac{\partial^2}{\partial y^2} + \frac{\partial^2}{\partial z^2}\right)\psi\left(x, y, z\right) + V\psi\left(x, y, z\right) = E\psi\left(x, y, z\right) \qquad (1.1)$$

Here $\psi(x, y, z)$ - is the wavefunction (eigenfunction) of the particle, $V$ - is the confinement potential, and $E$ - is the energy (eigenvalue) of the particle. Next, we need to define the boundary mesh region for the mesh regions used in our calculations are presented in Figure 1. The geometrical parameters used for the calculations in the current article are presented in the Table 1.

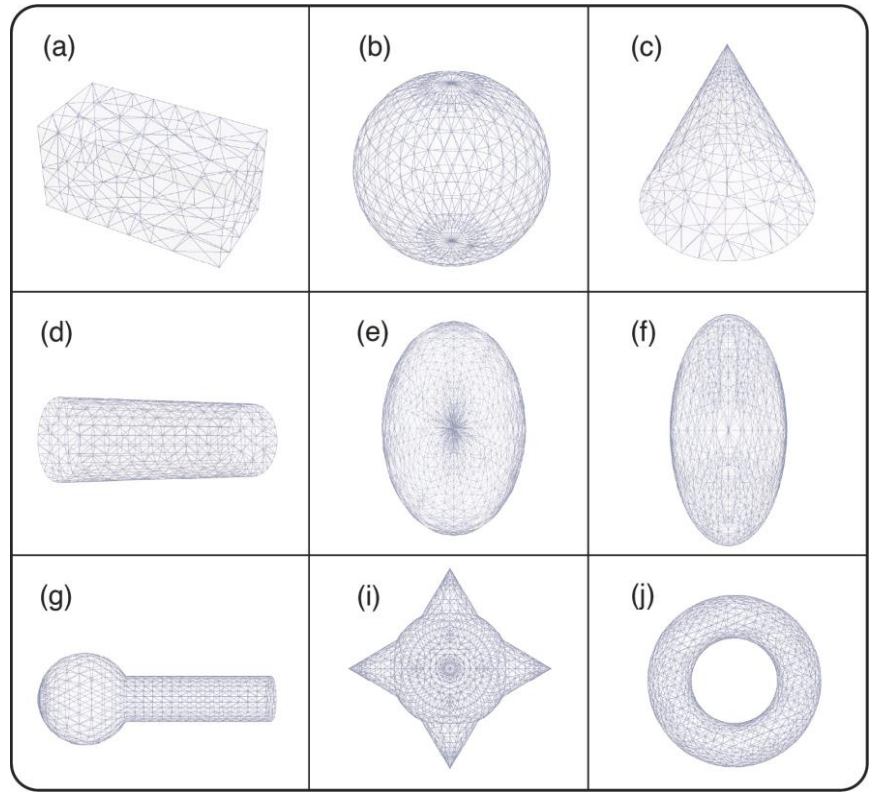

**Figure 1.** The mesh regions are used for the calculations. (a) mesh for a rectangular QD, (b) mesh for a spherical QD, (c) mesh for a conical QD, (d) mesh for a cylindrical QD, (e) mesh for an ellipsoid QD, (f) mesh for an ellipsoid of revolution (spheroid), (g) mesh for a nanotadpole, (i) mesh for a nanostar, (j) mesh for the quantum ring.

**Table 1.** Geometrical parameters of structures in the calculations.

| The Structure | Values of Geometrical Parameters (nm) |
|---|---|
| Rectangular QD | $L_x = 40, L_y = 20, L_z = 20$ |
| Spherical QD | $R = 10$ |
| Cylindrical QD | $R = 5, H = 30$ |
| Ellipsoidal QD | $a_x = 30, b_y = 10, c_z = 20,$ |
| Spheroidal QD | $a = 30, b = 20, c = 20$ |
| QR | $R_{inner} = 10, R_{outer} = 20$ |
| Conical QD | $R = 12, H = 20$ |
| Nanotadpole | $R_{head} = 12, R_{tail} = 6, H_{tail} = 20$ |
| Nanostar | $R_{sphere} = 10, R_{cone} = 10, H_{cone} = 7.5$ |

The final step of the FEM is the statement of the boundary conditions which is usually defined by the potential $V(x, y, z)$. In most of our examples, we are going to take into account the infinite well model where the probability density turns to zero outside of the structure $\left|\psi\left(outside\right)\right|^2 = 0$. This model corresponds to the potential of the form:

$$V(x, y, z) = \begin{cases} V(inside) = 0 \\ V(outside) = \infty \end{cases} \tag{1.2}$$

This implies that the particle is confined in the QD and cannot come outside of the structure.

After following these steps the FEM divides the mesh domain into finite elements and solves the equation. During these processes, the maximal size of these elements can be changed in general the smaller the elements higher the accuracy. In Mathematica, the size of these elements is defined by the property called MaxCellMeasure. To check the accuracy of the solution we can minimize the energy of the particle by varying the MaxCellMeasure parameter. The numerical error is also dependent on the geometrical parameters of the system. Another method for checking the accuracy of the method is to compare the eigenvalues obtained by FEM to the analytically solvable cases. The analysis of the ground state energy dependent on the MaxCellMeasure and the comparison of the first 25 eigenstates obtained by FEM to the analytically solvable QD models are presented in the 3.1 Accuracy and Computational time subsection of results and discussions.

Next, let us consider one of the methods for considering structures that contain different materials like core/shell QDs or multilayer QD. In such a system we have to take into account the change of the effective mass, dielectric permittivity, and potential depth and the effects caused by the diffusion. The $V(x, y, z)$ will have to change to a more complex form like a piecewise Woods-Saxon potential. For the sake of showcasing both the effective mass and the dielectric constant anisotropy we will consider a system with a hydrogen-like impurity at the center that has the following Hamiltonian:

$$-\frac{\hbar^2}{2m^*}\left(\frac{\partial^2}{\partial x^2} + \frac{\partial^2}{\partial y^2} + \frac{\partial^2}{\partial z^2}\right)\psi\left(x, y, z\right) + V\psi\left(x, y, z\right) + \frac{2e^2}{\varepsilon_0 \sqrt{(\vec{r} - r_0)^2}} = E\psi\left(x, y, z\right) \tag{1.3}$$

Where $r = \sqrt{x^2 + y^2 + z^2}$ - is the radius vector for the electron. For modeling the band structure of a spherical CdSe/CdS core/shell structure we can use:

$$V(r) = \begin{cases} V(r) = \left(V_0^{CdS} - V_0^{CdSe}\right) - \dfrac{V_0^{CdS} - V_0^{CdSe}}{1 + \exp\left[\left(r - R_{core}\right)/\alpha\right]}, & r < R_{core}. \\[4mm] V(r) = V_0^{CdS} - \dfrac{V_0^{CdSe}}{1 + \exp\left[\left(r - R_{core}\right)/\alpha\right]}, & r \geq R_{core}. \\[4mm] V(r) = \infty, & r \geq R_{shell}. \end{cases} \tag{1.4}$$

Here the $V_0^{CdSe} = 4.8\,eV$, $V_0^{CdS} = 4.9\,eV$ - is the position of the conduction band minimum for CdSe, CdS respectively, $R_{core}$ - is the core radius in our case CdSe region, $R_{shell}$ is the shell radius in our case CdS region, $\alpha$ - is a transition smoothness parameter which varies with the degree of diffusion. You can see the potential form in Figure 2.

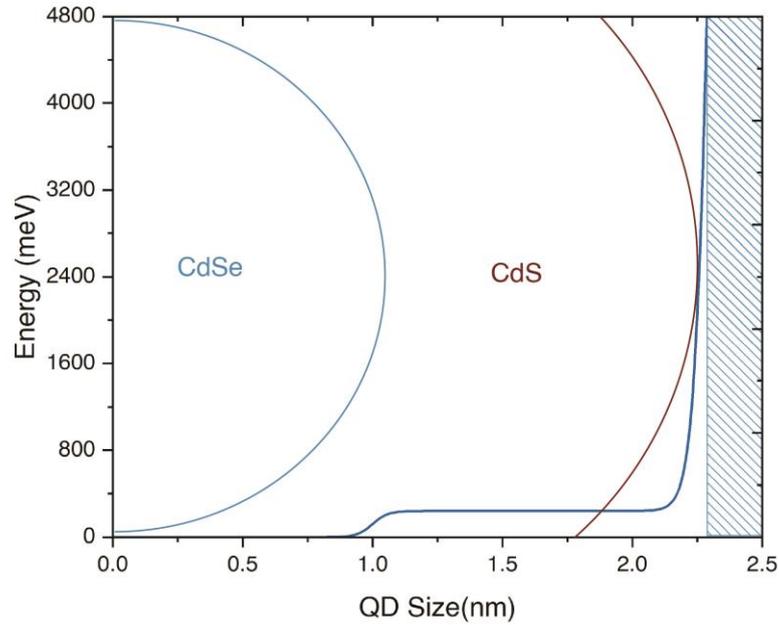

**Figure 2.** Piecewise Woods-Saxon Potential plotted for the following parameters: $R_{core} = 1nm$, $R_{shell} = 1.3nm$, $a_{core} = 0.03$.

The effective mass and dielectric permittivity can be defined by a standard Woods-Saxon potential:

$$m^*(r) = m^*_{CdS} - \frac{m^*_{CdS} - m^*_{CdSe}}{1 + \exp\left[\dfrac{r - R_{core}}{\alpha}\right]}$$ (1.5)

Here $m^*_{CdSe} = 0.112 \cdot m_0$, $m^*_{CdS} = 0.25 \cdot m_0$ - are the electron effective masses in the respective materials.

$$\varepsilon_0(r) = \varepsilon_{CdS} - \frac{\varepsilon_{CdSe} - \varepsilon_{CdS}}{1 + \exp\left[\dfrac{r - R_{core}}{\alpha}\right]}$$ (1.6)

Here $\varepsilon_{CdSe} = 9.29$, $\varepsilon_{CdS} = 8.28$ - are the electron effective masses in the respective materials. Lastly, it is important to mention that because the FEM sorts the states by increasing order of the energy we use the number of the state $n = 0, 1, ..., \infty$ which must not be confused with the quantum number $n$. Meaning $n$ is the ordering number.

The calculations were performed on a computer with the specifications presented in table 2. For a computer with a different parameters the computation time will be different. Other than that the results should be the same. Moreover the GPUs have little to no effect on FEM calculations we have brought the information for the sake of completeness.

**Table 2.** Geometrical parameters of structures in the calculations.

| Part name | Model |
|---|---|
| CPU | AMD Ryzen Thresadripper 3990X |
| RAM | Kingston 4x32GB 3600MHz |
| GPU | Double NVIDIA GeForce RTX 3090 |

### 3. Results & Discussions

#### 3.1 Acuracy and Computational time

Before proceeding to the investigation of the energy spectrum we should first showcase the limits of FEM's accuracy. It can be done in several ways. The most direct method for doing it is comparing the analytically obtainable energy values to the results obtained numerically. In general, we can assume that the numerical error increases for states with higher energy. The deviation from the exact analytical solution in the current article is defined as $d = 1 - E_{FEM}/E_{Analytic}$ its minimum value is 0 (exact solution). The deviations for the rectangular, spherical, and cylindrical QDs for the first twenty-six excited states are shown in Figure 3. As we can see the solution with the least deviation is obtained for the rectangular QD Figure3(a). And the solution with the most deviation and unpredictable behavior corresponds to the cylindrical QD. However, it is important to note that the deviation even in the worst-case scenario is less than 5%.

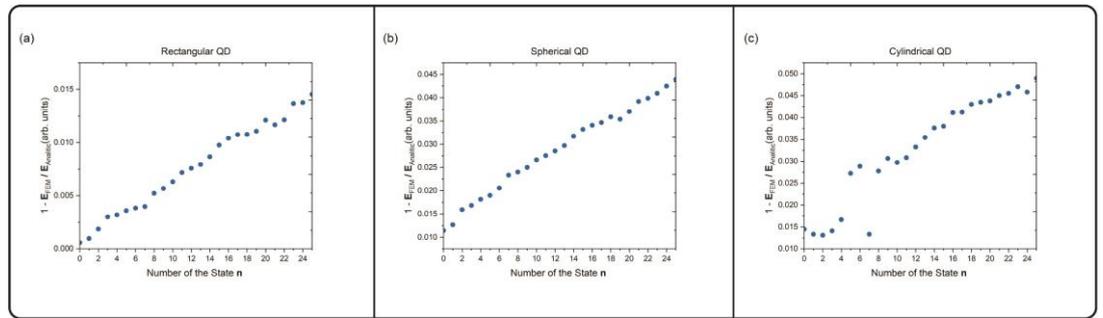

**Figure 3.** The deviation of the analytically obtained electron energy value from the numerical values for the first 26 states in (a) Rectangular, (b) spherical, and (c) cylindrical QDs. The comparison was carried out for the $MaxCellMeasure = 0.01$

For the QDs with non-trivial geometry, the accuracy of the FEM highly depends on how well the element mesh approximates the structure (region). Some geometrical parameters like extreme angles (near 180° or near 0°) can cause large discretization errors. Decreasing the size of the elements will increase the accuracy but cost more computational time. So it is important to find an optimal value for this parameter. To that end, we have calculated the ground state's energy $E_{ground}$ dependence on the MaxCellMeasure parameter. The FEM generally overestimates the energy values. We can assume that the lower the ground state energy the higher the accuracy of the solution. Thus in Figure 4, the dependence of the ground state energy on the MaxCellMeasure parameter is presented for the rectangular (a), spherical (b), cylindrical (c), ellipsoidal (d), spheroidal (e), and conical (g) QDs as well as QR (f), nanotadpoles (h), and nanostars (i). The parameter was varied in the range of $0.001 - 0.1$, we can see that universally the lowest value for the $E_{ground}$ is at the $MaxCellMeasure \approx 0.001$ which was to be expected. The most extreme decrease in energy is present for QRs and ellipsoidal QDs meaning that their geometries are the hardest to approximate with tetrahedral mesh elements used in Mathematica. However, it is visible that for most cases the energy decrease is in the order of 0.1meV which is negligible for most applications. So larger mesh elements can MaxCellMeasure ≈ 0.01 be comfortably used. However, in the case of the ellipsoidal QD and the QR, we can see that the energy difference is in the order of 1meV which cannot be neglected. The reverse side of the coin is the computation time which increases with the decrease of the mesh element size. The time for obtaining the first ten eigenvectors and eigenvalues for the abovementioned QDs is presented in Figure 5. The results again solidify the fact that for spheroidal QD decreasing the size of the mesh elements is a poor choice because the increase in computation time is the highest for it however the accuracy decrease is negligible.

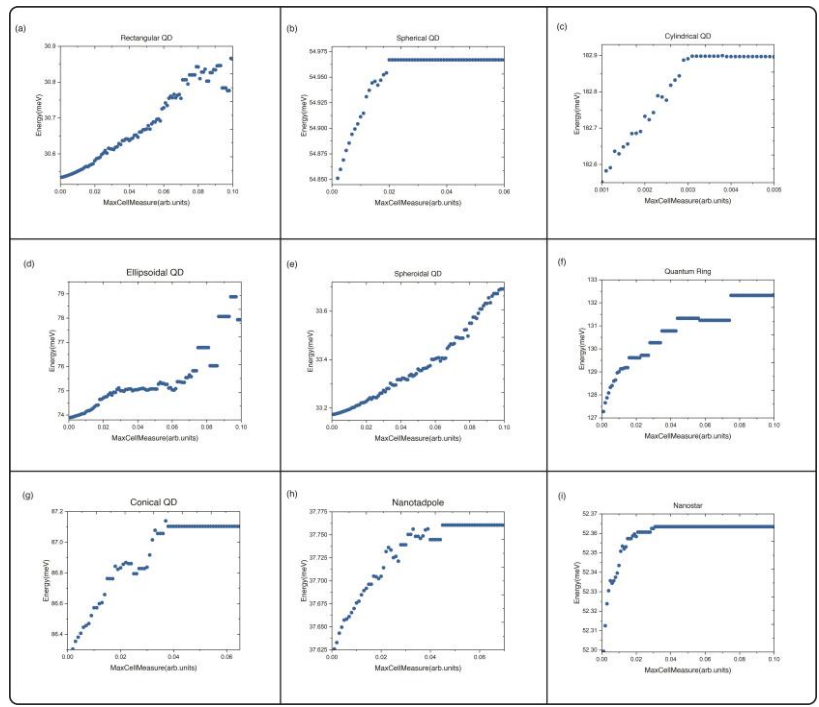

**Figure 4.** The dependence of the electron energy on the MaxCellMeasure parameter for (a) rectangular QDs, (b) spherical QDs, (c) cylindrical, (d) ellipsoidal QDs, (e) spheroidal, (f) QRs, (g) conical, (i) nanotadpoles QDs, (j) nanostars.

The results for the QRs showcase the relatively low time of 6s for the MaxCellMeasure 0.001 this makes the decrease of the MaxCellMeasure worthwhile. Lastly, although the computation cost for the ellipsoidal QD is the second highest at 46s the accuracy increase cannot be neglected and the lowest possible value should be taken for the best results. It is important to note that the discretization errors can be mitigated by the construction of custom meshes that use hexahedral elements or maybe even hybrid meshes consisting of hexahedral and tetrahedral elements that take into account the symmetry of the QDs more accurately. However, we believe that is beyond the scope of this paper.

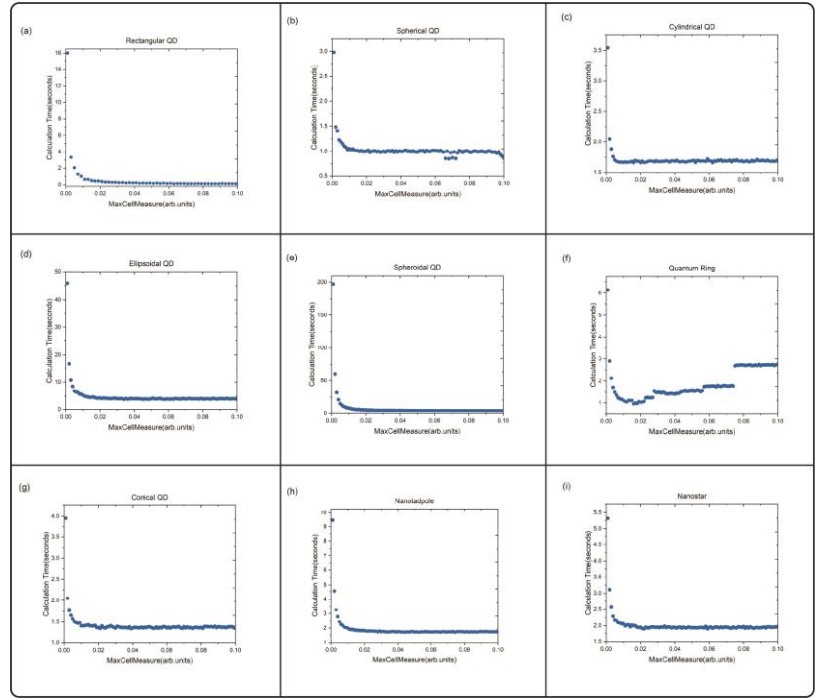

**Figure 5.** The dependence of computation time of the first 10 states on the `MaxCellMeasure` parameter for (a) rectangular QDs, (b) spherical QDs, (c) cylindrical, (d) ellipsoidal QDs, (e) spheroidal, (f) QRs, (g) conical, (i) nanotadpoles QDs, (j) nanostars.

*3.2 Electronic Energy and Wavefunctions of Semiconductor QDs*

The energy of the particle localized in a QD strongly depends on the size-quantization effects. This effect is what makes the spectrum of the QDs discrete. The size quantization effect depends on the material parameters of the structure, the size of the structure, and finally the shape of the structure. The size quantization not only shapes the energetic spectrum but also the wave function of the confined particle. As we know the wave functions of the state differ in size shape and orientation. The size quantization affects all of these features. And the changes in the probability density cause changes in the difference between the energy levels, number of degenerate states, etc. That is why understanding the energetic spectrum of QD requires a parallel analysis of the probability density.

Let us start from the simplest case an electron confined in the spherical QD the energy spectrum is shown in Figure 6(a) and the probability densities for the first four states are shown in Figure 7. As we can see the ground state has spherical symmetry and has no orbital nodes. The excited states probability densities have one orbital node and are oriented in three different directions. Because of the spherical symmetry, each direction is equivalent to the other, and states that are defined by the same number of orbital nodes that only differ in orientation are energetically degenerate. The degeneracy of the energy spectrum is consistent with the well known analytical solutions.

The situation changes in the case of a rectangular QD with geometrical parameters similar to the ones taken in our calculation $L_x = 40nm$, $L_y = 20nm$, $L_z = 20nm$. As we can see here the y and z directions have the same amount of size quantization. However, the size quantization is weak in the x directions. The degeneracy remains for wave functions oriented towards the y and z direction and this is reflected both by the probability densities (Figure 8) and the energy spectrum (Figure 6(b)). If the probability density is oriented in the x direction, the resulting state has lower energy. This effect is tangible to the point that the wavefunction with two orbital nodes oriented in the x direction has lower energy than wavefunctions with one orbital node oriented in the y and z directions.

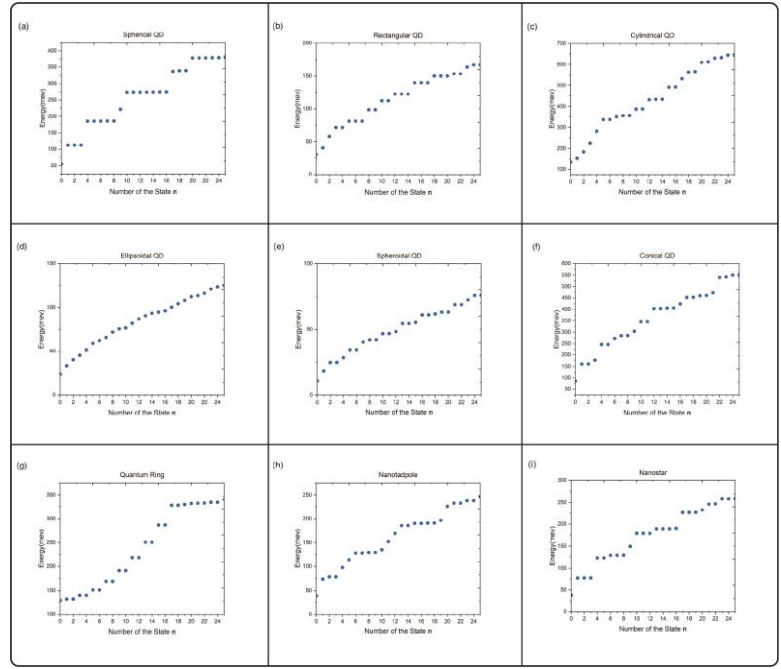

**Figure 6.** The values of energy for the electron's ground state and the first twenty-five excited states for (a) rectangular QDs, (b) spherical QDs, (c) cylindrical, (d) ellipsoidal QDs, (e) spheroidal, (f) QRs, (g) conical, (i) nanotadpoles QDs, (j) nanostars.

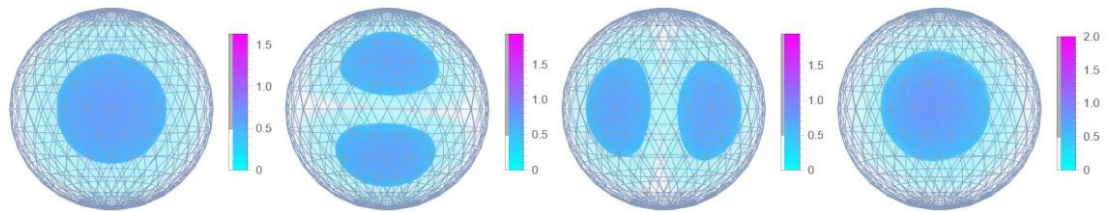

**Figure 7.** The probability density of the ground state and first three excited states of an electron confined in a spherical GaAs QD. The mesh domain used for the calculation is overlayed on the probability density.

This effect of the reordering of energy levels caused by the weaker size quantization in one direction is much more pronounced in the case of cylindrical QDs. The energy spectrum (Figure 6(c)) shows a non-degenerate behavior up to the sixth state and this is displayed in the probability density (Figure 9) where for the first five states the number of the nodes increases with each state. And the sixth and seventh states are the first states that are are not oriented in the z-direction and have one orbital node. The subsequent behavior of the energetic spectrum can be explained similarly.

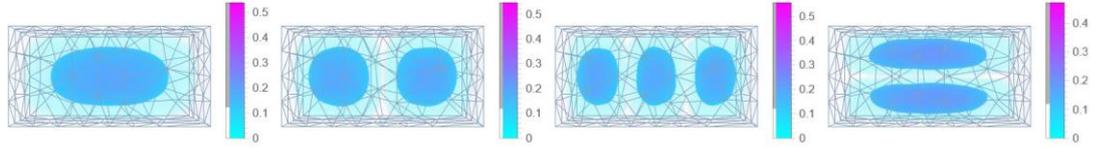

**Figure 8.** The probability density of the ground state and first three excited states of an electron confined in a rectangular GaAs QD. The mesh domain used for the calculation is overlaid on the probability density.

The first structure with the non-trivial geometry that we are going to present is the ellipsoidal QD with the following geometrical parameters $a_x = 30\,nm$, $b_y = 10\,nm$, $c_z = 20\,nm$.

All three half-axis have different values. This means that each direction has a different size quantization, which leads to the elimination of energetic degeneracy.

This is visible by the fact that the fourth state's wave function does not have one node and is not oriented in the y direction. Instead, the state with the one node wavefunction oriented towards the y-axis corresponds to the seventh state.

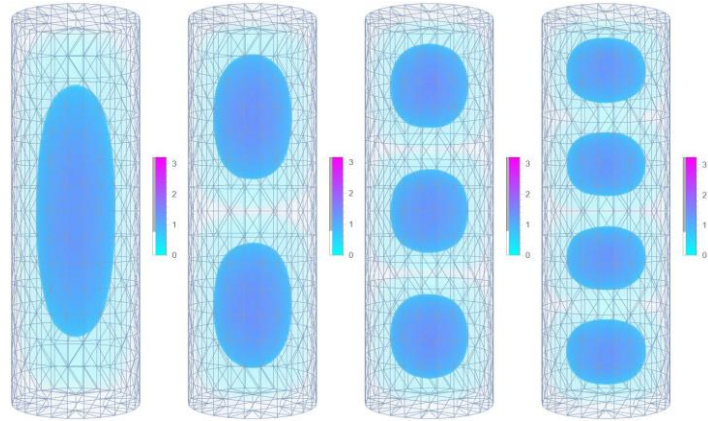

**Figure 9.** The probability density of the ground state and first three excited states of an electron confined in a cylindrical GaAs QD. The mesh domain used for the calculation is overlaid on the probability density.

We know that an ellipsoid becomes a sphere if all of the half-axis are equal to each other, and becomes a spheroid when two of the half-axis are equal. As such we can expect that an electron confined in a spheroidal QD will have a hybrid behavior between an ellipsoid and a sphere. The expected behavior is shown in Figure 10. The ground state has elliptic symmetry and the first three excited states all have one node similar to the spherical case. However, the difference arises from the fact that the wavefunction of the first excited state is oriented in the z-direction and subsequently has weaker size quantization causing the first excited state to lose its degeneracy. The energy spectrum continues this behavior to the higher states. It is important to note that for each case the ratio of the inequal direction to the equal directions plays a crucial role and the spectrum will reorder itself depending on the geometrical parameters.

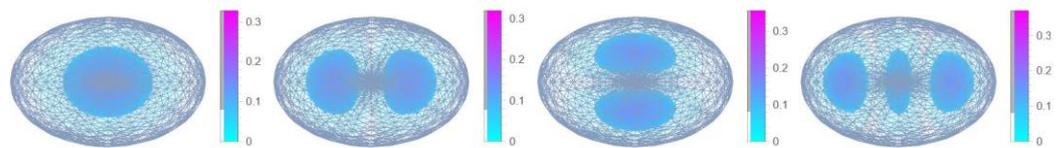

**Figure 9.** The probability density of the ground state and first three excited states of an electron confined in an ellipsoidal GaAs QD. The mesh domain used for the calculation is overlaid on the probability density.

This effect is caused by the directional non-equivalence and can be observed in any structure that has initial asymmetries. An example of such a structure is conical QDs (Figure 6(f) & Figure 11). The first 2 excited states have the same energies and wavefunctions only differing in their orientation. The third excited state is oriented towards the z-direction and has a higher energetic value because the electron is localized near the peak. Similar behavior can be observed in the energetic spectrum for the higher levels.

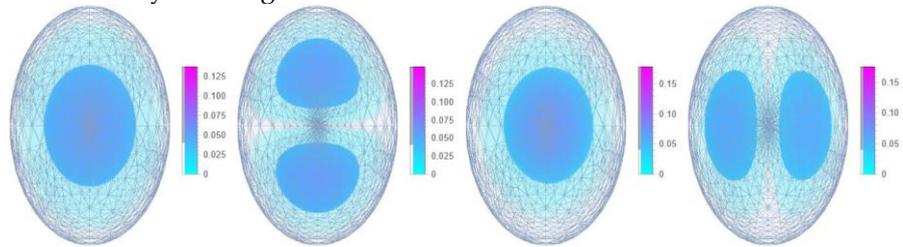

**Figure 10.** The probability density of the ground state and first three excited states of an electron confined in a spheroidal GaAs QD. The mesh domain used for the calculation is overlayed on the probability density.

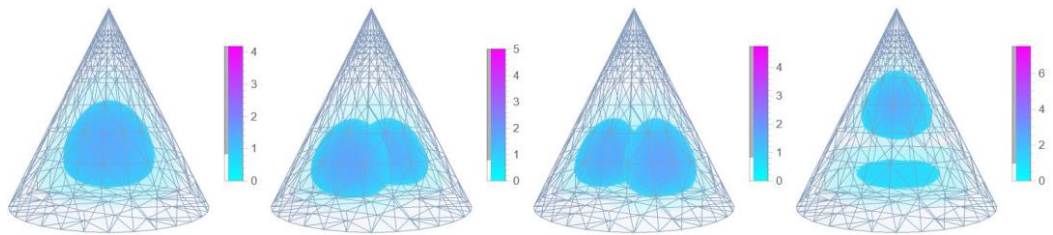

**Figure 11.** The probability density of the ground state and first three excited states of an electron confined in a conical GaAs QD. The mesh domain used for the calculation is overlayed on the probability density.

The reverse picture can be observed for a nanotadpole. Here we have a structure that is comprised of two regions a head region approximated with a sphere and a tail region approximated with a cylinder merged to the head in the z-direction. This asymmetry in the z-direction leads to the lowering of the size quantization and a highly unusual shape for the wave function. However, the shape for the second and third excited states almost matches the spherical QD. This means that the electronic probability density only "seeps through" to the tail region in the case when the wave function is to some extent oriented in the z-direction (Figure 12). The energetic spectrum (Figure 6(h)) shows the effects of this "seep through" with the partial elimination of degenerate levels.

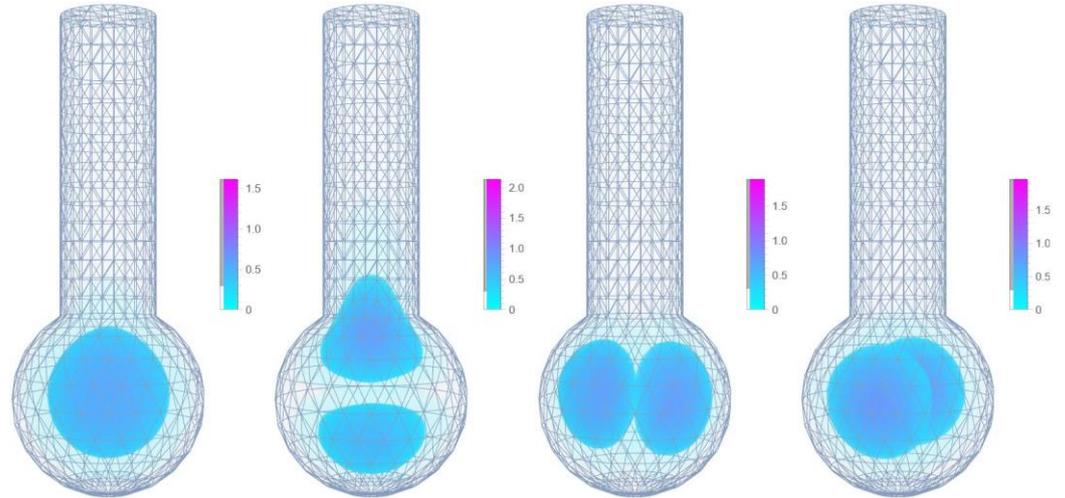

**Figure 12.** The probability density of the ground state and first three excited states of an electron confined in a conical GaAs nanotadpole. The mesh domain used for the calculation is overlayed on the probability density.

However, the directional non-equivalence is partially eliminated if you have a symmetrical structure like a nanostar which is symmetrical towards the three axes. At the first glance you can see in Figure 13 the wavefunctions are extremely similar to the spherical case. However, the additional freedom provided by the points leads to the transformation of the wave functions in all directions. For the ground state, the difference is slightly visible, however, for the excited state, the most noticeable contribution is the change in the orientation as you can see the first three excited states are no longer aligned strictly to the x, y and z-axis.

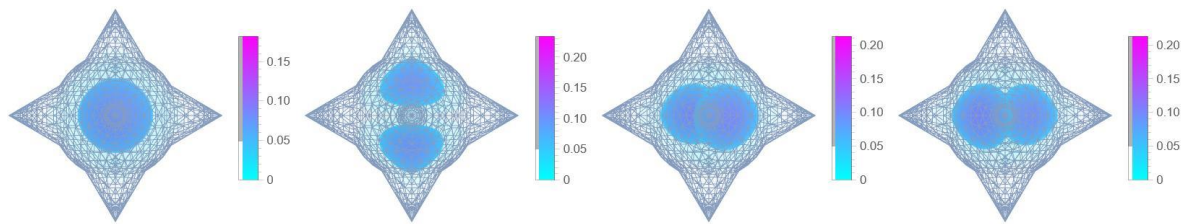

**Figure 13.** The probability density of the ground state and first three excited states of an electron confined in a GaAs nanostar. The mesh domain used for the calculation is overlayed on the probability density.

Let us close this sub-section with the QRs. Here the structure's symmetry leads to the most unique ground state probability density that we have discussed. It has a ring-like shape and is distributed uniformly in the structure creating a "ring of probability density". The excited states have more shapes and more possible orientations compared to the sphere.

In the first twenty-five states, there are four states with wavefunctions that have one orbital node. However, only two of them degenerate in energy in the second and third excited states (Figure 14). The other two correspond to the twentieth and twenty-second excited states, the former's probability densities are comprised of two probability density "rings" outgoing from the center (similar to the rings of Saturn), and the latter's two probability density "rings" are stacked on each other. These shape and orientation changes contribute to the extreme energy difference between the first two and last two states with one node wavefunctions. If we look at the energy spectrum Figure 6(g) we can see that the states are doubly degenerate or near degenerate until the sixteenth state.

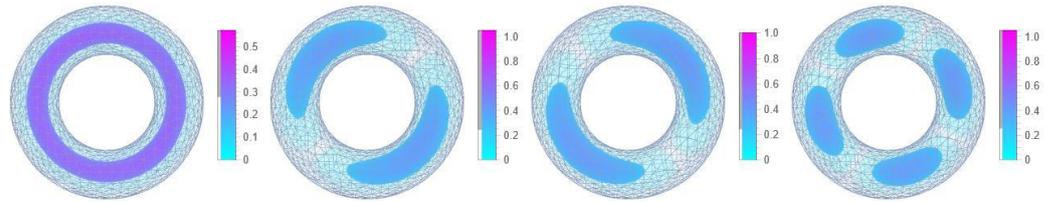

**Figure 14.** The probability density of the ground state and first three excited states of an electron confined in a GaAs QR. The mesh domain used for the calculation is overlayed on the probability density.

These results showcase the way by which we can control the overlap between electronic and excitonic states and the energy spectrum, and by extension most of the optoelectronic properties of QD.

### 3.3 Properties of Semiconductor Core/Shell QDs

In this subsection, we are going to investigate the localization probability for electrons and holes confined in CdSe/CdS core-shell spherical QD with an impurity in the center.

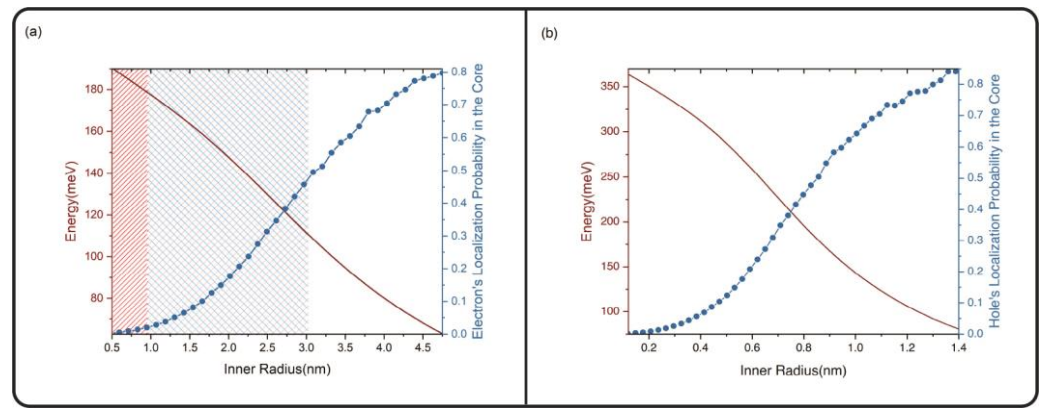

**Figure 15.** The segmented blue line is the probability density of the particle's ground state localized in the core region, the solid red line is the ground state energy of the particle depending on the inner core size for an electron(a) and a hole(b). In Figure (a) you can see three regions the region that has a red patterns represents the core radii where the localization probability of the hole in the core is less than 50%. The second region represented by the red and blue pattern shows the core radii where the hole's localization probability in the core is larger than 50% but the electron's localization probability is lower than 50%. The third region does not have pattern, and shows the core radii where the both hole's and electron's localization probability in the core are larger than 50%.

This type of structure can be approximated by a potential of the form (1.4) that is presented in Figure 2. Generally, core/shell structures are grown with materials that have a bigger potential depth in the core $V_0^{CdSe}$ so for medium and large core sizes $R_{core}$, the particles are localized in the core. Let us first discuss the electron localization probability of an electron in Figure 15(a). If the electron's energy is noticeably lower than the difference between the confinement potentials of the two materials $V_{e0}^{CdS} - V_{e0}^{CdSe} \approx 100 meV$. The electron is mostly localized in the core. However, by making the core radius smaller the size quantization in the core becomes stronger and the electron's probability density "seeps through" to the shell region more and more as ground state energy rises. The electron localization probability density's dependence on the core radius $R_{core}$ is presented in Figure 15(a) with the segmented blue line, the ground state energy is presented in the same graph with the red solid line. You can see that around the point $R_{core} = 4.74 nm$, about 80% of the electron probability density is localized in the core. As the size of the core decreases electron probability "seeps through" more and more to the shell, until at $R_{core} = 2.2 nm$ only ~ 20% is in the core. It is important to note that the $V_{e0}^{CdS} - V_{e0}^{CdSe}$ depends on the growing methods and for different values the probability density will behave differently. This is also visible for the hole. The potential difference for the hole $V_{h0}^{CdS} - V_{h0}^{CdSe}$ is twice as large $V_{h0}^{CdS} - V_{h0}^{CdSe} \approx 200 meV$. The hole's localization probability in the core (Figure 15(b)) reaches about 85% at the core size $R_{core} = 1.4 nm$. This is drastically different from the electron which means that for core size values around 2.2 nm the overlap between the electron and hole wave functions decrease to the point that the structure can be considered to have a quasi type-II band structure. These kinds of structures have a plethora of advantages namely the exciton lifetime increases drastically. Quasi type-II structures are actively investigated experimentally. For example in [76], the authors have investigated the temperature dependence of the spectral properties such as the band gap, bandwidth, and fluorescence intensity of CdSe/CdS dot-in-rod nanocrystals. Quasi type-II structures were synthesized with core sizes of $R_{core} = 2.3 nm$. The values for quasi-type-II structure core size obtained by our calculations are very close to the experimental values, which attests to the quality of our chosen model.

Overall, we can say that the FEM can not only be used to model one material QDs but core/shell, core/shell/shell, or dot in bulk structures successfully. Even allowing us to obtain structures with quasi type-II band alignment.

## 5. Conclusions

Summarizing the results obtained in the scope of the current paper we can say that the FEM can be used to model QDs with various trivial and non-trivial geometries as well as multi-material QDs. To recap subsection 3.1: the deviation between the FEM and the analytically obtainable energy values of the first twenty-six states was less than $d \approx 0.05$ for a $MaxCellMeasure = 0.01$. The dependence of the ground state energy $E_{ground}$ on the element size $MaxCellMeasure$ showed a significant correlation for QRs and ellipsoidal QDs (in the order of 1 meV). This shows that these structures are hard to approximate by Mathematica's default discretization algorithm. This is also reflected by the fact that the computation cost for obtaining the first ten wavefunctions and energies for ellipsoidal QD at the $MaxCellMeasure = 0.001$ is the second highest at 46s. However, despite this relatively large timeframe, the impact on accuracy cannot be neglected. The situation is much more positive for the QR the time for completing the calculation with the same parameters is about $6s$. In the next subsection of results and discussions 3.2, the electronic energy spectrum and probability densities were investigated for rectangular, spherical, cylindrical, ellipsoidal, spheroidal, conical QDs, QRs, nanotadpoles QDs, and nanostars. The dependence of the energy spectrum and wavefunction orientation on the size quantization effect's directional non-equivalence was discussed in detail. The effect of various asymmetries of structures was obtained and discussed. In the last subsection of results and discussions, we have obtained the particle's localization probability in the core of the CdSe/CdS structure for the electron and the hole. It was found that it is possible to obtain a structure with quasi type-II band alignment using the model chosen in our calculations. And the core size values correspond to the experimentally obtained quasi type-II structures. Overall, we can say that the FEM is a flexible tool for modeling QDs with various parameters. And the accuracy largely depends on the chosen potential models and the constructed mesh.